\documentclass[twoside]{article}
\usepackage{graphicx}
\usepackage{epsfig}
\usepackage{fancyhdr}
\usepackage{multicol}
\usepackage{styl-ap}

\begin{document}
\title{The First Stars}
\author{Daniel J. Whalen\work{1}}
\workplace{McWilliams Fellow, Department of Physics, Carnegie Mellon University, Pittsburgh, PA 15213}
\mainauthor{dwhalen@lanl.gov}
\maketitle

\begin{abstract}%

Pop III stars are the key to the character of primeval galaxies, the first heavy elements, the onset of 
cosmological reionization, and the seeds of supermassive black holes.  Unfortunately, in spite of their 
increasing sophistication, numerical models of Pop III star formation cannot yet predict the masses of 
the first stars. Because they also lie at the edge of the observable universe, individual Pop III stars will 
remain beyond the reach of observatories for decades to come, and so their properties are unknown.  
However, it will soon be possible to constrain their masses by direct detection of their supernovae, and 
by reconciling their nucleosynthetic yields to the chemical abundances measured in ancient metal-poor 
stars in the Galactic halo, some of which may bear the ashes of the first stars.  Here, I review the 
state of the art in numerical simulations of primordial stars and attempts to directly and indirectly 
constrain their properties observationally.

\end{abstract}


\begin{multicols}{2}
\section{The Simulation Frontier}

Unlike with star formation in the Galaxy today, there is little disagreement over the initial conditions 
of star formation in the primordial universe.  The original numerical simulations suggested that Pop 
III stars formed in small pregalactic structures known as cosmological minihalos at $z \sim$ 20 - 30, 
or $\sim$ 200 Myr after the Big Bang (Bromm, Coppi \& Larson, 1999,2001; Abel, Bryan \& Norman, 
2000, 2002; Nakamura \& Umemura, 2001).  These models predicted that the stars formed in 
isolation, one per halo, and that they were 100 - 500 M$_{\odot}$.  Pop III stars profoundly 
transformed the halos that gave birth to them, expelling their baryons in supersonic ionized flows 
and later exploding as supernovae (e.g. Whalen, Abel \& Norman, 2004; Kitayama \& Yoshida 2005; 
Whalen \& Norman 2008; Whalen et al. 2008a).  Radiation fronts from these stars also engulfed 
nearby halos, either promoting or suppressing star formation in them, thereby regulating the rise of 
the first stellar populations (Shapiro et al. 2004; Iliev et al. 2005; Susa \& Umemura 2006; Whalen et 
al., 2008b,2010; Hasegawa et al. 2009; Susa et al. 2009).

The original estimates of Pop III stellar masses were not obtained by modeling the actual formation 
and evolution of the stars.  They were derived by comparing infall rates at the center of the halo at 
early stages of collapse to Kelvin-Helmholz contraction times to place upper limits on the final mass 
of the star. Later simulation campaigns in the same vein revealed a much broader range of final 
masses for Pop III stars, 30 - 300 M$_{\odot}$ (O'Shea \& Norman, 2007), and that they could form 
as binaries in a fraction of the halos (Turk et al., 2009).  Heroic numerical efforts have only recently 
achieved the formation of a hydrostatic protostar at the center of the halo (Yoshida et al., 2008) and 
the collapse of the central flow into an accretion disk (Stacy et al., 2010; Clark et al., 2011; Smith et 
al., 2011; Greif et al., 2011,2012). 

\begin{myfigure}
\centerline{\resizebox{70mm}{!}{\includegraphics{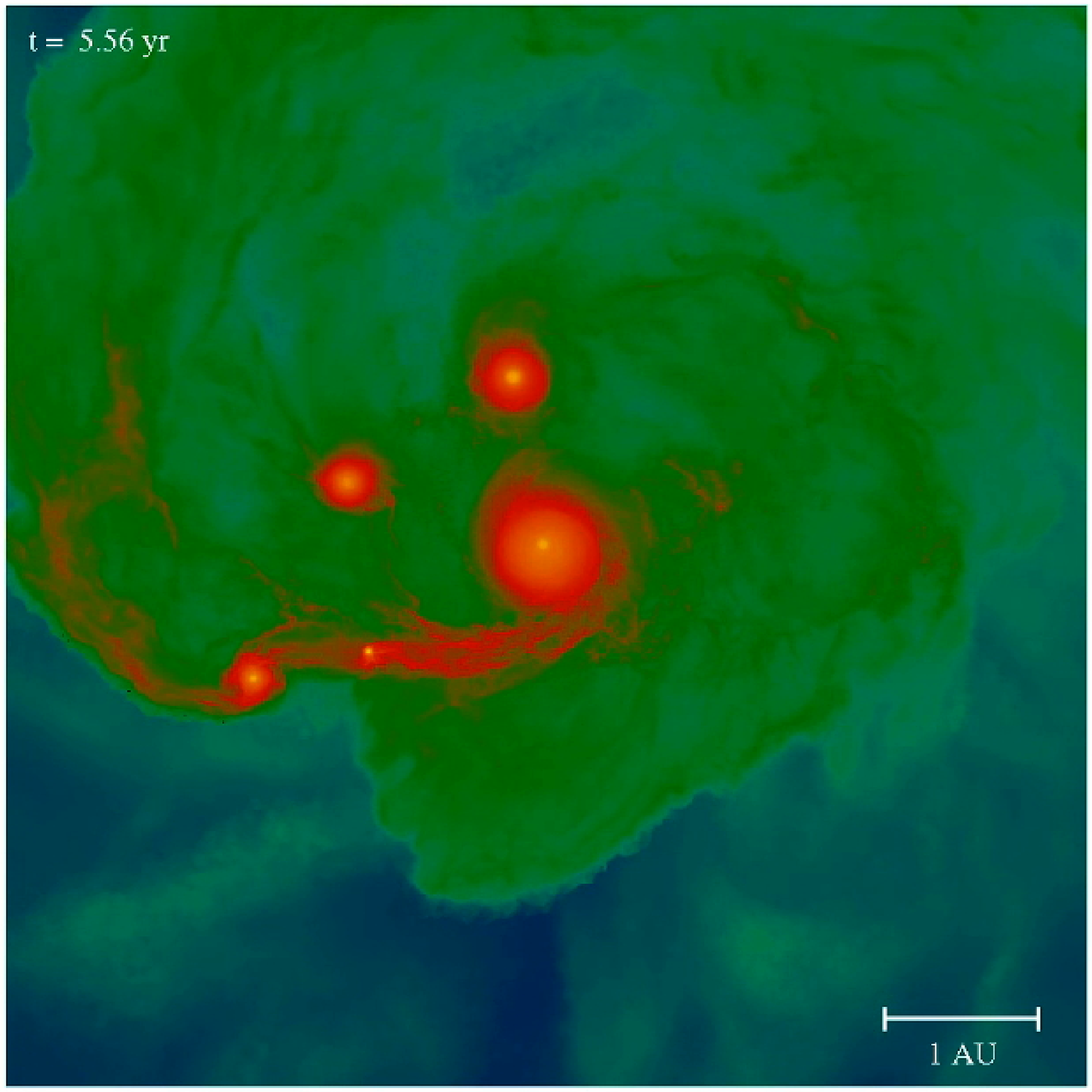}}}
\caption{The formation and fragmentation of a Pop III protostellar disk in the Arepo code (Greif et al., 
2012).}
\label{fig:popIIIdisk}
\end{myfigure}

In particular, the disk calculations indicate that they are unstable to fragmentation, raising the 
possibility that Pop III stars may have only been tens of solar masses, not hundreds, and that they 
may have formed in small swarms of up to a dozen at the centers of primeval halos.  Computer 
models of ionizing UV breakout in the final stages of Pop III protostellar disks have also found that 
the I-front of the nascent star exits the disk in bipolar outflows that terminate accretion onto the star 
and mostly evaporate the disk by the time the star reaches $\sim$ 40 M$_{\odot}$ (Hosokawa et al., 
2011).  This result reinforces the sentiments of some in the community that while the Pop III IMF 
was top-heavy, primordial stars may only have been 10 - 40 M$_{\odot}$.

\subsection{High-Mass or Low-Mass Pop III Stars?}

In spite of their increasing sophistication, these simulations should be taken to be very preliminary 
for several reasons.  First, Pop III accretion disks form in smoothed-particle hydrodynamics (SPH) 
models but not in adaptive mesh refinement (AMR) simulations, although the AMR models have 
not evolved the collapse of the halo to the times achieved by SPH calculations.  This raises the 
question of whether one technique better captures the transport of angular momentum out of the 
center of the cloud than the other, and whether accretion is ultimately spherical or through a disk.  
Second, the stability of the disk itself remains an open question because although the simulations 
can now fully resolve the disk they do not yet incorporate all of its relevant physics.  In particular, 
they lack high-order radiation transport, which regulates the thermal state of the disk and its 
tendency to fragment.  Furthermore, the role of primordial magnetic fields in the formation and 
evolution of the disk is not well understood (Turk et al., 2012; Widrow et al., 2012).  

The use of sink particles to represent disk fragments in the original SPH simulations of Pop III 
protostellar disk formation called into question the longevity of the fragments.  Once they are 
created in the simulations they are never destroyed, unlike real fragments that could be torn 
apart by gravitational torques and viscous forces (Norman, 2010).  More recent moving mesh 
simulations performed with the Arepo code that do not rely on sink particles find that the 
fragments persist but only evolve the disk for 10 - 20 yr (Greif et al., 2012).  Perhaps most 
importantly, no simulation has followed the disks for more than a few centuries, far short of the 
time required to assemble a massive star.  Thus, it remains unclear if the fragments in the disk 
remain distinct objects or merge with the largest one at the center, building it up into a very 
massive star over time through protracted, clumpy accretion.

\subsection{Accretion Cutoff and the Final Masses of Pop III Stars}

This latter point directly impacts estimates of final masses for Pop III stars inferred from numerical 
simulations that attempt to model how ionizing UV from the star reverses infall and evaporates the 
accretion disk.  At the heart of such models is a simple recipe for the evolution of the protostar that 
provides a prescription for its radius and luminosity as a function of time and acquired mass.  The 
Hosokawa et al. (2011) 2D calculations take the growth of the protostar to be relatively steady, in 
which case it contracts and settles onto the main sequence at $\sim 30$ M$_{\odot}$.  At this point 
the star becomes extremely luminous in ionizing UV radiation that halts accretion onto the star in a 
few hundred kyr at a final mass of $\sim$ 40 M$_{\odot}$.  If accretion instead turns out to be 
clumpy, the protostar could remain puffy and cool and reach much larger masses before burning 
off the disk.  

A finer point is that all current accretion cutoff simulations evolve both radiation and hydrodynamics
on the Courant time, a practice which is known to lead to serious inaccuracies in I-front propagation 
in density gradients (Whalen \& Norman, 2006).  Such coarse time steps may result in premature 
I-front breakout and accretion cutoff, and hence underestimates of the final mass of the star.  Three
dimensional simulations with more accurate radiation--matter coupling schemes, both steady and 
clumpy accretion scenarios, more realistic prescriptions for protostellar evolution based on 
nucleosynthesis codes such as KEPLER (Weaver et al., 1978; Woosley et al., 2002) and a variety of 
halo environments may better constrain the Pop III IMF.  However, in judging the power of such 
simulations to model the masses of the first stars, it should be remembered that no simulations 
realistically bridge the gap in time between the formation and fragmentation of a protostellar disk 
and its photoevaporation up to a Myr later.  We note in passing that fragments can also stop 
accreting if they are ejected from the disk by 3-body gravitational effects (Greif et al., 2011; Johnson 
\& Khochfar, 2011).  These fragments could become very low-mass Pop III stars ($\sim$ 1 M$_{
\odot}$); if so, some of them may live today.

\section{Constraining the Pop III IMF with Stellar Archaeology}

\begin{figure*}
\begin{center}
\begin{tabular}{cc}
\epsfig{file=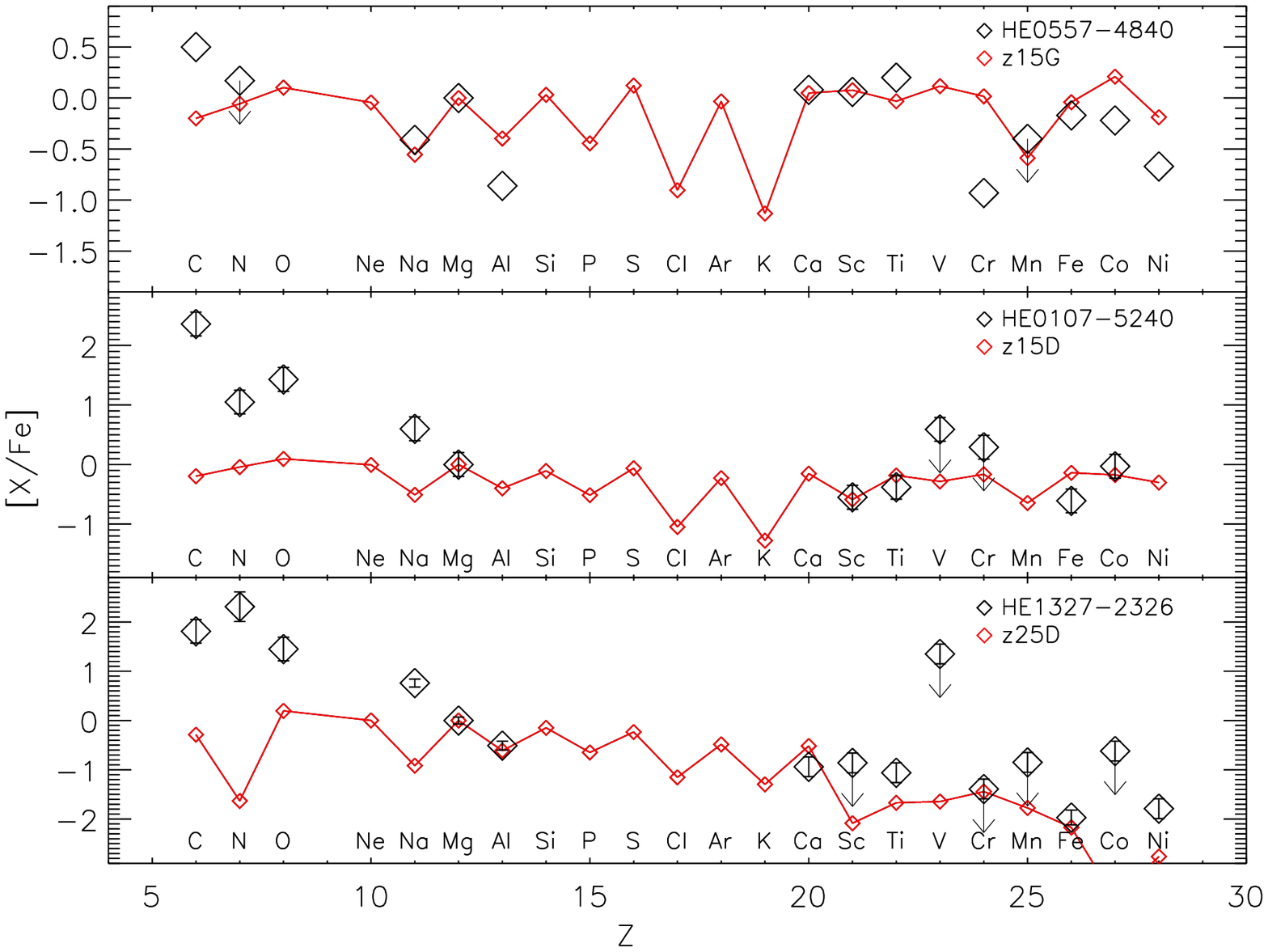,width=0.40\linewidth,clip=} & 
\epsfig{file=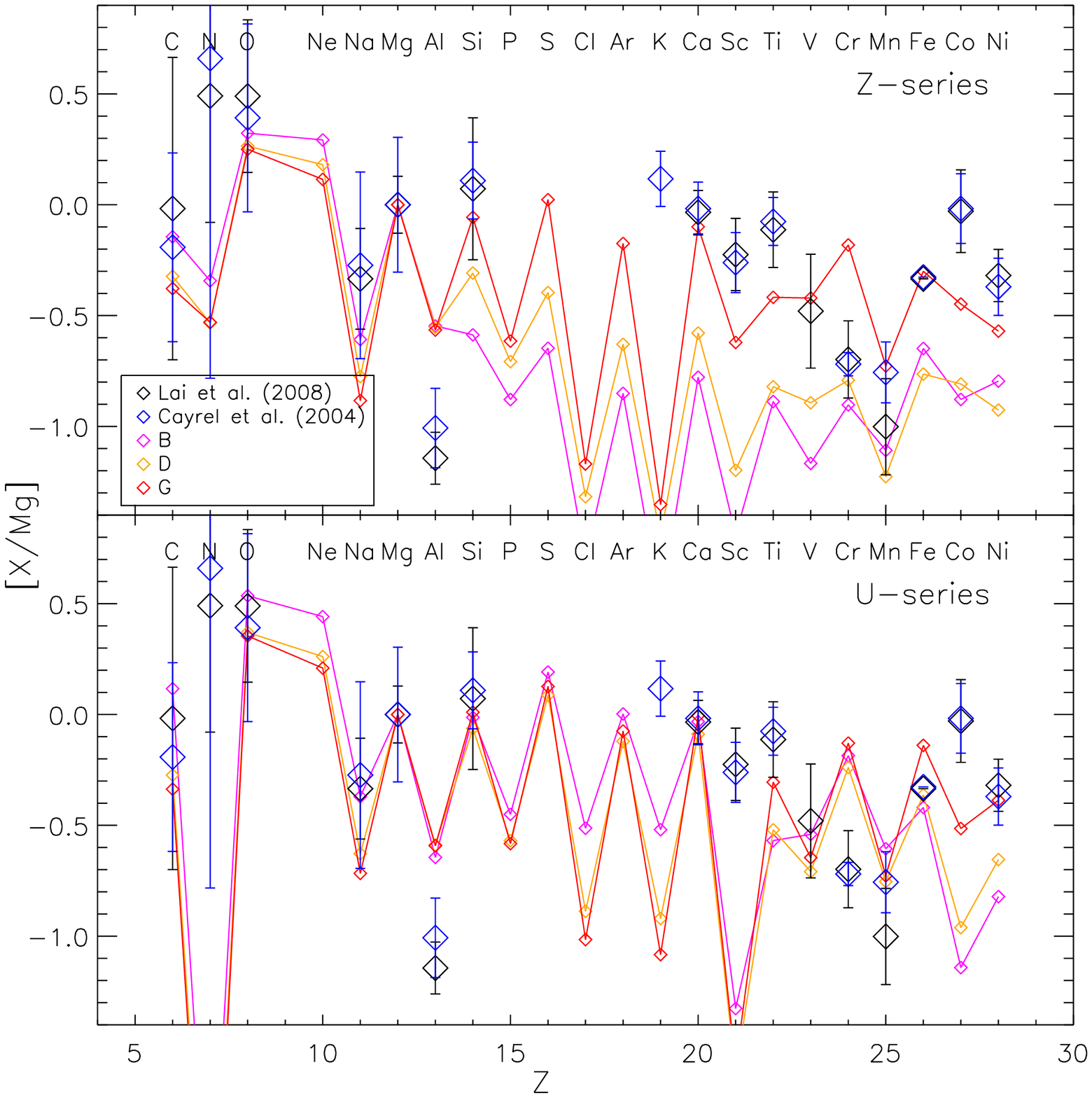,width=0.40\linewidth,clip=} \\
\end{tabular}
\end{center}
\caption{Comparing Pop III SN yields to the chemical abundances of three of the most metal-poor 
stars (left panel) and the extremely metal-poor (EMP) stars in the Cayrel et al. (2004) and Lai 
et al. (2008) surveys (right panel).  In the left panel, the abundances in HE0557-4840 agree well 
with the yields from SN model z15G in the Joggerst et al. (2010) study.  In the right panel we 
show that higher explosion energy rotating $Z=0$ stars reproduce EMP abundances well.  The 
existence of 15 M$_{\odot}$ Pop III stars is required to produce this good agreement with 
observations.}
\label{fig:yields}
\end{figure*}

Unfortunately, because they lie at the edge of the observable universe, individual Pop III stars 
will remain beyond the reach of direct detection for decades to come, even with their enormous 
luminosities (Schaerer 2002) and the advent of the next generation of near infrared (NIR) 
observatories such as the James Webb Space Telescope (JWST) and the Thirty-Meter Telescope 
(TMT).  However, there have been attempts to indirectly constrain the masses of Pop III stars by 
comparing the cumulative elemental yield of their supernovae to the fossil chemical abundances 
found in ancient metal-poor stars in the Galactic halo, some of which may be second-generation 
stars.  Stellar evolution models indicate that 15 - 40 M$_{\odot}$ primordial stars died in core
collapse (CC) supernovae (SNe) and that 40 - 140 M$_{\odot}$ stars collapsed to black holes, 
perhaps with violent pulsational mass loss prior to death (Heger \& Woosley, 2002).  Pop III stars 
between 140 and 260 M$_{\odot}$ can die in pair-instability (PI) SNe, with energies up to 100 
times those of Type Ia SNe that completely unbind the star and leave no compact remnant 
(Chatzopoulos \& Wheeler 2012 have recently discovered that rotating Pop III stars down to 65 
M$_{\odot}$ can also die as PI SNe).  These explosions were the first great nucleosynthetic 
engines of the universe, expelling up to half the mass of the progenitor in heavy elements into 
the early IGM.  Primordial stars above 260 M$_{\odot}$ collapsed directly to black holes, with no 
mass loss.

Joggerst et al. (2010) recently calculated the chemical imprint of low-mass Pop III SNe on later 
generations of stars by modeling mixing and fallback onto the central black hole in 15 - 40 M$_
{\odot}$ Pop III core collapse explosions with the CASTRO AMR code.  As shown in Figure 
\ref{fig:yields}, a simple power-law IMF average of the elemental yields of these explosions is in 
good agreement with the fossil abundances in a sample of 130 extremely metal poor stars with 
$Z < 10^{-4}$ $Z_{\odot}$ (Cayrel et al., 2004; Lai et al., 2008).  Although these results suggest 
that low-mass Pop III stars shouldered the bulk of the chemical enrichment of the early IGM, 40 
- 60 M$_{\odot}$ hypernova explosions, whose energies are intermediate to those of CC and PI 
SNe, may also have contributed metals at high redshifts (Iwamoto et al. 2005).  

To date, the telltale odd-even nucleosynthetic signature of PI SNe has not been found in the fossil 
abundance record, leading some to assert that Pop III stars could not have been very massive. 
However, the odd-even effect may have been masked by observational bias in previous surveys 
(Karlsson et al., 2008).  Reconciling Pop III SN yields to the elemental patterns in metal-poor 
stars is still in its infancy for several reasons.  First, only small numbers of extremely metal-poor 
stars have been discovered to date, and larger sample sizes would better constrain early SN yields.  
Second, measurements of some elements in low-metallicity stars are challenging and in the past 
have been subject to systematic error.  Finally, there are many intervening hydrodynamical 
processes between the expulsion of the first metals and their uptake into new stars that are not 
yet understood.

\section{Finding the First Cosmic Explosions}

\begin{figure*}
\begin{center}
\begin{tabular}{cc}
\epsfig{file=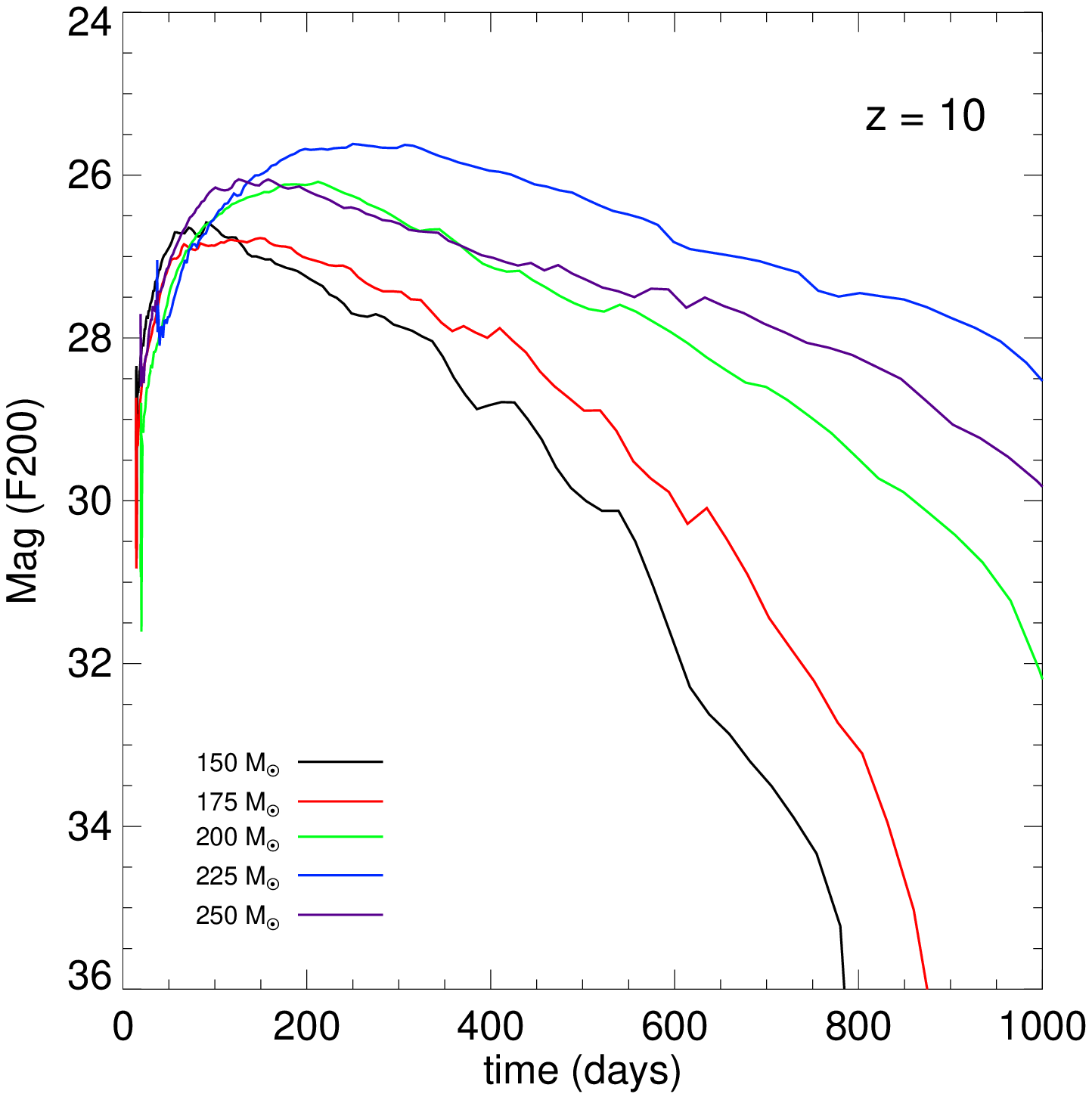,width=0.40\linewidth,clip=} & 
\epsfig{file=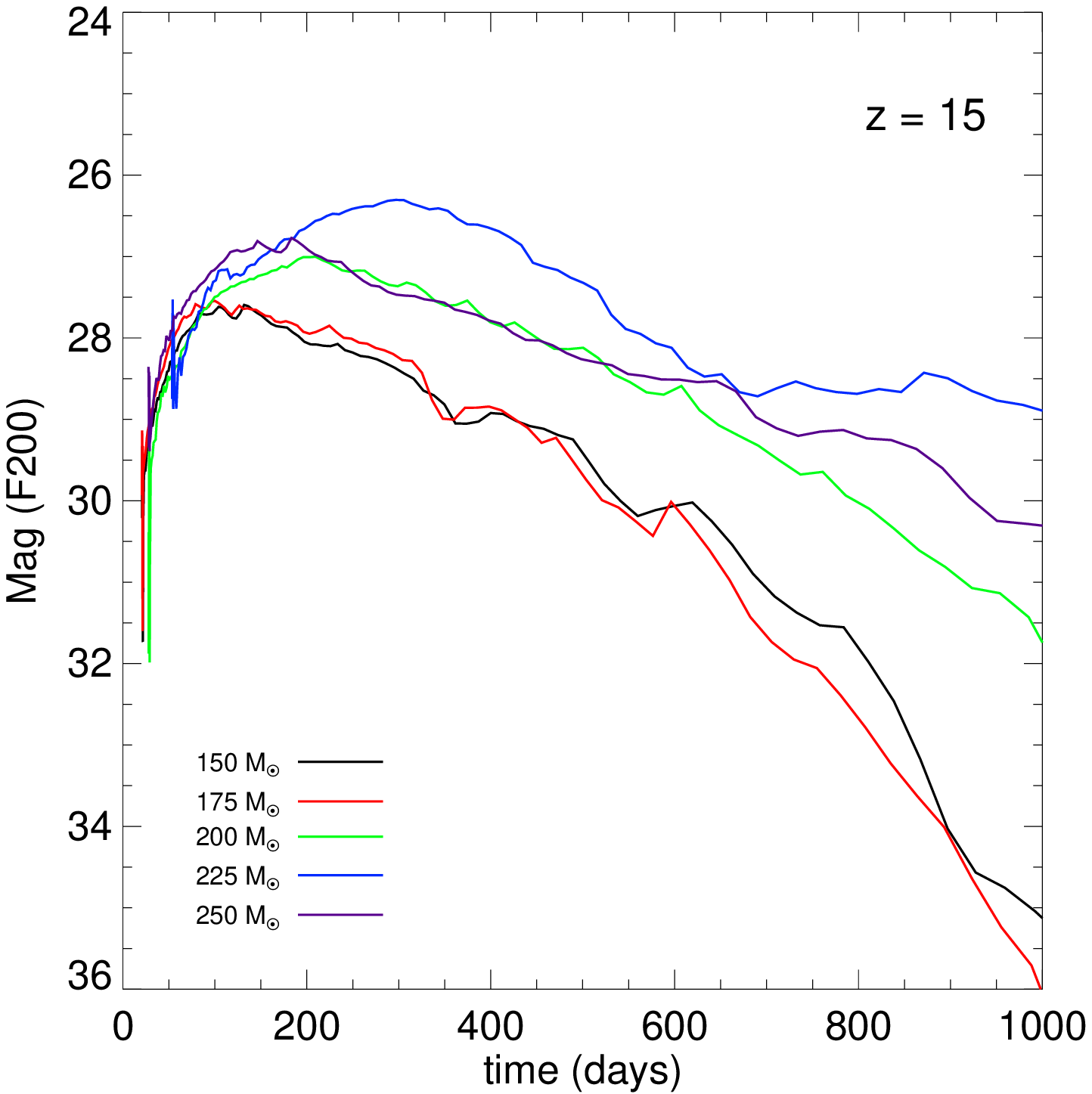,width=0.40\linewidth,clip=} \\
\epsfig{file=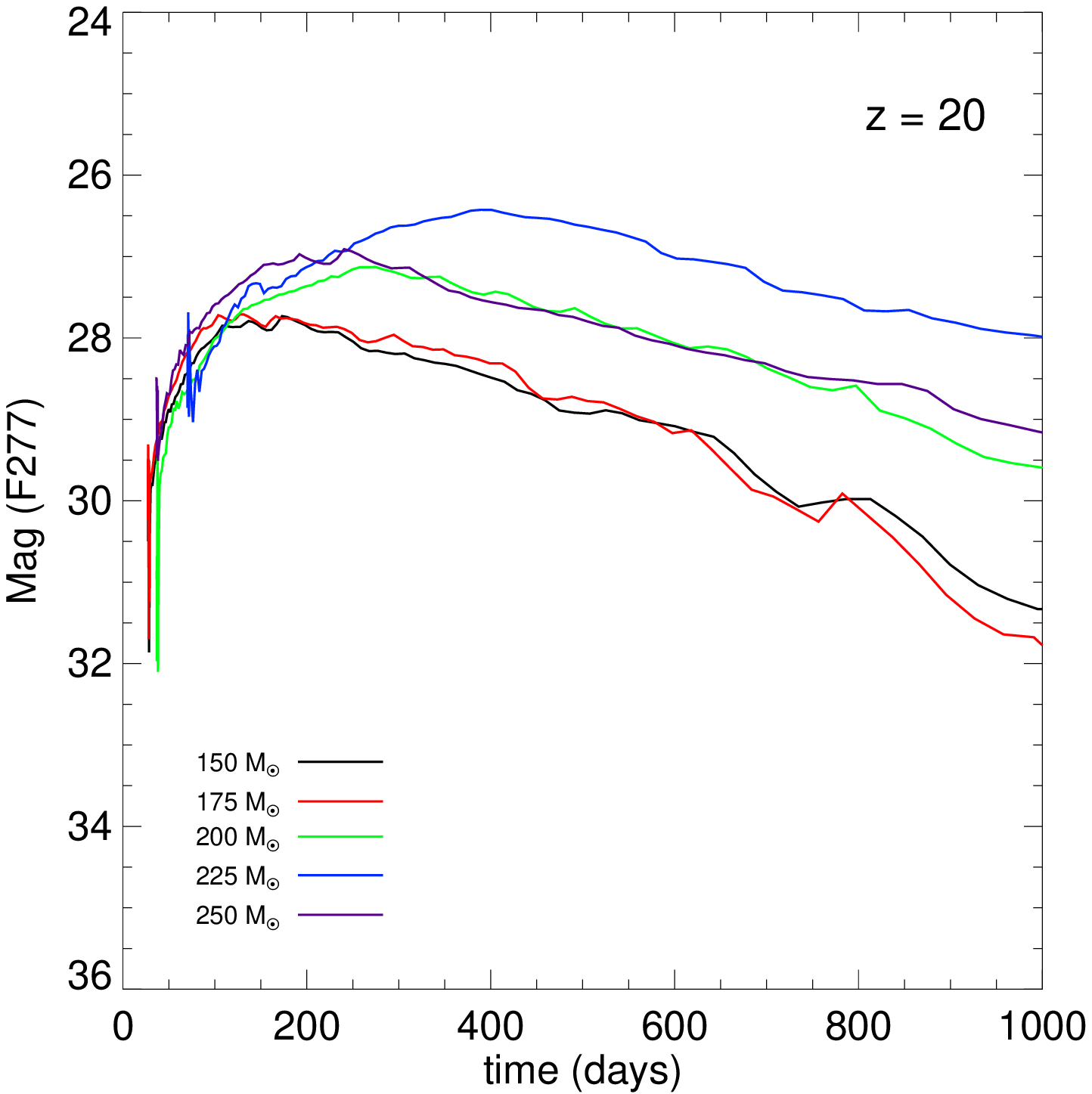,width=0.40\linewidth,clip=} &
\epsfig{file=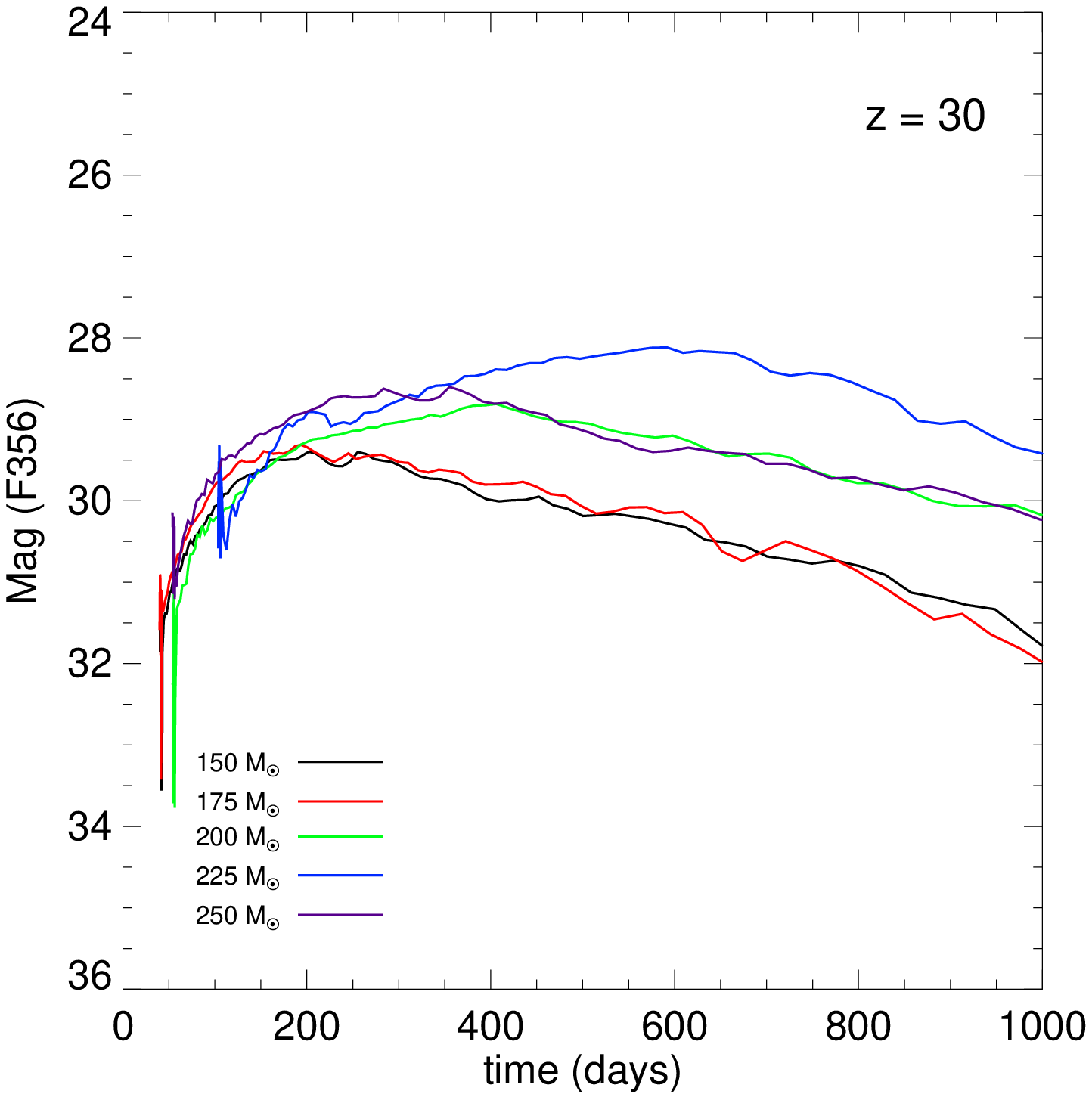,width=0.40\linewidth,clip=}
\end{tabular}
\end{center}
\caption{Pop III PI SN light curves for the JWST NIRCam. Clockwise from the upper left panel, 
the redshifts are $z =$ 10, 15, 20 and 30.  The optimum filter for each redshift is noted on the 
y-axis labels and the times on the x-axes are for the observer frame.  F200, F277 and F356 
are at 2.0, 2.77 and 3.56 $\mu$m, respectively.}
\label{fig:JWST_LC}
\end{figure*}

Detection of Pop III SNe would unambiguously probe the masses of primordial stars for the 
first time.  Since these explosions are 100,000 times brighter than either their progenitors or 
the primitive galaxies that host them, they could be found by JWST or the Wide-Field Infrared 
Survey Telescope (WFIRST). However, unlike the Type Ia SNe used to constrain cosmic 
acceleration, light from primeval supernovae must traverse the vast cosmic web of neutral 
hydrogen that filled the universe prior to the epoch of reionization.  Lyman absorption by this
hydrogen removes or scatters most of the light from ancient supernovae out of our line of 
sight, obscuring them.  

Whalen et al. (2012a,b,c) have calculated JWST NIR light curves for Pop III pair-instability SNe 
with the Los Alamos RAGE and SPECTRUM codes (Gittings et al., 2008; Frey et al., 2012), 
which are shown for $z =$ 10, 15, 20 and 30 in Figure \ref{fig:JWST_LC}.  These simulations
include radiation hydrodynamical calculations of the SN light curve and spectra in the local 
frame, cosmological redshifting, and Lyman absorption by intergalactic hydrogen.  JWST 
detection limits at 2 - 4 $\mu$m are AB magnitude 31 - 32, so it is clear that JWST will be able 
to detect the first cosmic explosions in the universe if they are PI SNe (and even perform 
spectrometry on them).  Even given JWST's very narrow fields of view at high redshifts, 
recent calculations indicate that at least a few PI SNe should be present in any given JWST 
survey (Hummel et al. 2012).  Also, because WFIRST detection limits will be AB magnitude 26.5
at 2.2 $\mu$m, it is clear from Figure \ref{fig:JWST_LC} that Pop III PI SNe will be visible to 
WFIRST out to $z \sim$ 15 - 20. Since it is an all-sky survey, and because this redshift range 
may favor the formation of very massive Pop III stars because of the rise of Lyman-Werner UV 
backgrounds (O'Shea \& Norman, 2008), WFIRST will detect much larger numbers of Pop III 
SNe.  

Could Pop III SNe be detected at later stages by other means?  Whalen et al., (2008b) found 
that most of the energy of Pop III CC SNe is eventually radiated away as H and He lines as the 
remnant sweeps up and shocks surrounding gas. At later epochs this energy would instead be 
lost to fine structure cooling by metals.  In both cases the emission is too dim, redshifted and 
drawn out over time to be detected by any upcoming instruments.  However, PI SNe deposit
up to half of their energy into the cosmic microwave background (CMB) by inverse Compton 
scattering at $z \sim$ 20 (Kitayama \& Yoshida, 2005; Whalen et al., 2008b) and could impose 
excess power on the CMB at small scales (Oh, Cooray \& Kamionkowski, 2003).  The resolution 
of current ground-based CMB telescopes such as the \textit{Atacama Cosmology Telescope} 
and \textit{South Pole Telescope} approaches that required to directly image Sunyaev-Zeldovich 
(SZ) fluctuations from individual Pop III PI SN remnants, so future observatories may detect 
them.  Unlike PI SNe, CC SNe deposit little of their energy into the CMB at $z \sim$ 20, and 
even less at lower redshifts because the density of CMB photons falls with cosmological 
expansion. 

The extreme NIR luminosities of primordial PI SNe could contribute to a NIR background 
excess, as has been suggested for Pop III stars themselves, i.e. Kashlinsky (2005).  New 
calculations reveal that enough synchrotron emission from CC SN remnants is redshifted 
into the 21 cm band above $z \sim$ 10 to be directly detected by the \textit{Square Kilometer 
Array} (\textit{SKA}) (Meiksen \& Whalen 2012).  Somewhat more energetic hypernovae could 
be detected by existing facilities such as the \textit{Extended Very-Large Array} ({\textit{eVLA}}) 
and \textit{eMERLIN}. PI SN remnants generally expand into ambient media that are too diffuse
to generate a detectable synchrotron signal.  Pop III SN event rates make it unlikely that they
will be found in absorption at 21 cm at $z >$ 10.     

The detection of the first cosmic explosions will be one of the most spectacular discoveries in 
extragalactic astronomy in the coming decade, opening our first observational window on the 
era of first light and the end of the cosmic Dark Ages at $z \sim$ 30.  They will unveil the nature 
of primordial stars and constrain scenarios for early cosmological reionization, the process 
whereby the universe was gradually transformed from a cold, dark, featureless void into the vast, 
hot, ionized expanse of galaxies we observe today.  At somewhat lower redshifts ($z \sim$ 10 - 
15), detections of Pop III supernovae will probe the era of primitive galaxy formation, marking the 
positions of nascent galaxies on the sky that might otherwise have eluded detection by JWST.  
Finally, finding the first supernovae could also reveal the masses of the seeds of the supermassive 
black holes lurking at the centers of massive galaxies today. 

\thanks

DJW was supported by the Bruce and Astrid McWilliams Center for Cosmology at Carnegie 
Mellon University.  Work at LANL was done under the auspices of the National Nuclear Security 
Administration of the U.S. Dept of Energy at Los Alamos National Laboratory under Contract No. 
DE-AC52-06NA25396. All numerical simulations were performed on Institutional Computing (IC) 
and Yellow network platforms at LANL (Conejo, Lobo and Yellowrail).

\bigskip
\bigskip
\noindent {\bf DISCUSSION}

\bigskip
\noindent {\bf Wolfgang Kundt:} You spoke of a mass range ($> 40$ M$_{\odot}$) for primordial 
stars for which they would collapse to black holes.  Where are the nearest of them today?

\bigskip
\noindent {\bf Daniel Whalen:} This is an excellent question.  We recently published a letter (Whalen 
\& Fryer 2012) in which we found that most 20 - 40 M$_{\odot}$ Pop III black holes would be ejected
from the cosmological halos that gave birth to them at velocities of 500 - 1000 km s$^{-1}$ by natal
kicks due to asymmetries in the core-collapse engine.  Such velocities are far above the escape 
velocity of any halo they would encounter for over a Hubble time, so there is a good chance that 
these black holes would be exiled to the voids between galaxies today.  Black holes above 40 M$_{
\odot}$ are unlikely to be born with kicks and remain in the halo, accreting and growing over cosmic 
time.  These black holes are much more likely to reside in the galaxies into which their host halos 
were taken, a few of which could become the supermassive black holes found in the SDSS quasars 
today.

\bigskip
\noindent {\bf Maurice Van Putten:} What fraction of the gas in a cosmological halo ends up in 
primordial stars?

\bigskip
\noindent {\bf Daniel Whalen:} It is currently thought that the minimum halo mass for forming a Pop 
III star is $\sim$ 10$^5$ M$_{\odot}$ and that 1 - 10 stars are formed with masses of 30 - 300 M$_
{\odot}$.  Thus, a conservative estimate is that 0.1 - 1\% of the baryons in the halo are converted
into stars, and that the rest are evicted from the halo by strong ionized flows over the life of the 
stars.

\end{multicols}

\begin{thebibliography}{99}

\bibitem{} Abel, T., Bryan, G. \& Norman, M.~L.:  2000, 540, 39.

\bibitem{} Abel, T., Bryan, G. \& Norman, M.~L.:  2002, Science, 295, 93.

\bibitem{} Bromm, V., Coppi, P.~S. and Larson, R.~B.:  1999, ApJL, 527, L5.

\bibitem{} Bromm, V., Coppi, P.~S. and Larson, R.~B.: 2002, ApJ, 564, 23.

\bibitem{} Cayrel, R. et al.:  2004, A\&A, 416, 1117.

\bibitem{} Clark, P.~C. et al.:  2011, Science, 331, 1040.

\bibitem{} Frey, L.~H. et al.:  2012, ApJS, submitted, arXiv:1203.5832.

\bibitem{} Gittings, M. et al.:  2008, Comp. Sci. \& Disc., 1, 015005.

\bibitem{} Greif, T.~H. et al.:  2011, ApJ, 737, 75.

\bibitem{} Greif, T.~H. et al.:  2012, MNRAS, 424, 399.

\bibitem{} Hasegawa, K. et~al.: 2009, MNRAS, 445

\bibitem{} Heger, A. \& Woosley, S.~E.:  2002, ApJ, 567, 532.

\bibitem{} Hosokawa, T. et al.:  2011, Science, 334, 1250.

\bibitem{} Hummel, J. et al.:  2012, ApJ, submitted, arXiv:1112.5207.

\bibitem{} Iliev, I. et al.:  2005, MNRAS, 361, 405

\bibitem{} Iwamoto, N. et al.:  2005, Science, 309, 451.

\bibitem{} Johnson, J.~L. \& Khochfar, S.:  2011, MNRAS, 413, 1184.

\bibitem{} Joggerst, C.~C. et al.:  2010, ApJ, 709, 11.

\bibitem{} Karlsson, T., Johnson, J.~L. \& Bromm, V.:  2008, ApJ, 679, 6.

\bibitem{} Kashlinsky, A. et~al.:  2005, Nature, 438, 45.

\bibitem{} Kitayama K. \& Yoshida, N.:  2005, ApJ, 630, 675

\bibitem{} Lai, D.~K. et al.:  2008, ApJ, 681, 1524.

\bibitem{} Meiksen, A. \& Whalen, D.~J.: 2012, MNRAS, submitted, arXiv:1209:1915.  

\bibitem{} Nakamura, F. \& Umemura, M.:  2001, ApJ, 548, 19.

\bibitem{} Norman, M.~L.:  2010, in \textit{First Stars and Galaxies: Challenges in the Coming 
Decade}, AIP Conf. Ser. 1294, 17.

\bibitem{} Oh, S.~P., Cooray, A. \& Kamionkowski, M.:  2003, MNRAS, 342, L20.

\bibitem{} O'Shea, B.~L. \& Norman, M.~L.:  2007, ApJ, 654, 66.

\bibitem{} O'Shea, B.~L. \& Norman, M.~L.:  2008, ApJ, 673, 14.

\bibitem{} Schaerer, D.:  2002, A\&A, 382, 28.

\bibitem{} Shapiro, P.~R. et~al.:  2004, MNRAS, 348, 753

\bibitem{} Smith, R.~J. et al.:  2011, MNRAS, 414, 3633.

\bibitem{} Stacy, A. et al.:  2010, MNRAS, 403, 45.

\bibitem{} Susa, H.  \& Umemura, M.:  2006, ApJL, 645, L93

\bibitem{} Susa, H. et al.:  2009, ApJ, 702, 480

\bibitem{} Turk, M.~J. et al.:  2009, Science, 325, 601.

\bibitem{} Turk, M.~J. et al.:  2012, ApJ, 745, 154.

\bibitem{} Weaver, T.~A., Zimmerman, G.~B. \& Woosley, S.~E.:  1978, ApJ, 225, 1021.

\bibitem{} Whalen, D.~J., Abel, T. \& Norman, M.~L.:  2004, ApJ, 610, 14.

\bibitem{} Whalen, D.~J. et al.:  2008a, ApJ, 682, 49.

\bibitem{} Whalen, D.~J. et al.:  2008b, ApJ, 679, 925.

\bibitem{} Whalen, D.~J. et al.:  2010, ApJ, 712, 101.

\bibitem{} Whalen, D.~J. \& Fryer, C.~L.:  2012, ApJL, ApJL, 756, L19

\bibitem{} Whalen, D.~J. \& Norman, M.~L.: 2006, ApJ, 162, 281.

\bibitem{} Whalen, D.~J. \& Norman, M.~L.: 2008, ApJ, 673, 664.

\bibitem{} Whalen, D.~J. et al.:  2012a, arXiv:1209.3457

\bibitem{} Whalen, D.~J. et al.:  2012b, arXiv:1209.5459

\bibitem{} Whalen, D.~J. et al.:  2012c, ApJ, submitted

\bibitem{} Widrow, L.~M. et al.:  2012, SSR, 166, 37.

\bibitem{} Woosley, S.~E., Heger, A. \& Weaver, T.~A.:  2002, Rev. Mod. Phys., 74, 1015.

\bibitem{} Yorke, H.  \& Welz, A.:  1996, A\&A, 315, 555.

\bibitem{} Yoshida, N., Omukai,  K \& Hernquist, L.:  2008, Science, 321, 669.

\end{thebibliography}
\end{document}